\documentclass [12 pt]{article}
\def\be{\begin{equation}}
\def\eea{\end{eqnarray}}
\def\bea{\begin{eqnarray}}
\def\ee{\end{equation}}

\usepackage[psamsfonts]{amssymb}
\usepackage{amsmath}

\author{M. Amooshahi$^{1}$ \footnote{amooshahi@sci.ui.ac.ir} and F.
Kheirandish $^{2}$ \footnote{fardin$_{-}$kh@phys.ui.ac.ir}
\\ $^{1}$ {\small Faculty of Science, University of Isfahan, Hezar Jarib Avenue,
Isfahan,Iran}\\
$^{2}$ {\small Faculty of Science, University of Isfahan, Hezar
Jarib Avenue, Isfahan,Iran}}
\title{Electromagnetic field quantization in an anisotropic magnetodielectric medium with spatial-temporal
dispersion}
\begin{document}
\maketitle
\begin{abstract}

\noindent

By modeling a linear, anisotropic and inhomogeneous
magnetodielectric medium with two independent set of harmonic
oscillators, electromagnetic field is quantized in such a medium.
The electric and magnetic polarizations of the medium are expressed
as linear combinations of the ladder operators describing the
magnetodielectric medium. The Maxwell and the constitutive equations
of the medium are obtained as the Heisenberg equations of the total
system. The electric and magnetic susceptibilities of the medium are
obtained in terms of the tensors coupling the medium with the
electromagnetic field. The explicit forms of the electromagnetic
field operators are obtained in terms of the ladder operators of the
medium.


\end{abstract}

\section{Introduction}
One  method to quantize the electromagnetic field in the presence
of an absorptive medium is known as the Green function method
\cite{1}-\cite{8}. In this method by adding the noise electric
and magnetic polarization densities to the classical constitutive
equations of the medium, these equations are taken as definitions
of electric and magnetic polarization operators. The noise
polarizations are related to two independent sets of bosonic
operators. Combination of Maxwell and the constitutive equations
in frequency domain, gives the electromagnetic field operators in
terms of the noise polarizations and classical Green tensor. The
commutation relations are imposed on the bosonic operators such
that the commutation relations between electromagnetic field
operators in the magnetodielectric medium become identical with
those in free space. Another interesting quantization scheme of
electromagnetic field in the presence of an absorptive dielectric
medium is known as the damped polarization model which is based
on the Hopfield model of a dielectric\cite{9}, where the
polarization of the dielectric is represented by a damped quantum
field \cite{10}. In the damped polarization model
\cite{11}-\cite{13} , the electric polarization of the medium is
represented by a quantum field and absorptivity character of the
medium is described by interaction between the polarization with
a heat bath containing a continua of harmonic oscillators. In
this method a canonical quantization is formulated for
electromagnetic field and the medium. The dielectric function of
the medium is obtained in terms of coupling function of the heat
bath and the electric polarization, such that it satisfies the
Kramers-Kronig relations \cite{14}. Recently Raabe and et al
\cite{14.1} have represented a unified method to quantize
electromagnetic field in the presence of an arbitrary linear
medium based on a general nonlocal conductivity tensor and using
a single set of appropriate bosonic operators. This formalism
recovers and generalizes the previous  quantization schemes for
diverse classes of linear media. In particular the  quantization
of electromagnetic field in the presence of a magnetodielectric
medium is a limiting case for weakly spatially dispersive medium
in this scheme. In the present work, we generalize our approach
\cite{15,16} to quantizing electromagnetic field in an anisotropic
magnetodielectric medium with spatially and temporarily
dispersive property. In this case, the electric and magnetic
polarizations are dependent on the macroscopic electric and
magnetic fields inside the medium in a non local way with respect
to both position and time. The electric and magnetic polarization
densities of the medium are defined as linear combinations of the
ladder operators of the medium. The coefficients of these linear
expansions are coupling tensors which couple the medium with
electromagnetic field. The electric and magnetic susceptibility
tensors of the medium are obtained in terms of the coupling
tensors. By using a Hamiltonian in which the electric and
magnetic polarizations minimally couple to the displacement and
magnetic fields respectively, both Maxwell and
 the constitutive equations of the medium can be obtained as the
Heisenberg equations of the total system. Finally, using the Laplace
and Fourier transformations,
 we obtain the space-time dependence of electromagnetic
field operators in terms of the annihilation and creation operators
of the oscillators modeling the medium.
\section{Quantization scheme}

In order to quantize electromagnetic field in the presence of an
anisotropic magnetodielectric medium, we  enter the medium
directly in the process of quantization modeling it by two
independent set of harmonic oscillators which we call them $E$
and $M$ quantum fields  \cite{15}. The $E$ and $M$ fields
describe polarizability and magnetizability of the medium,
respectively. This means that in our approach the electric and
magnetic polarization densities of the medium are defined
respectively as linear combinations of the ladder operators of the
$E$ and $M$ quantum fields. We use the Coloumb gauge in this
quantization scheme and do the quantization in unbounded space and
in the absence of external charges. Generalization of the
quantization inside a cavity with a definite volume and with a
known boundary conditions and or in the presence of external
charges is straightforwardly \cite{16}.

Applying the Coloumb gauge, the quantum vector potential can be
expanded in terms of plane waves as
\begin{equation} \label{d1}
\vec{A}(\vec{r},t)=\int d^3 \vec{k} \sum_{\lambda=1}^2
\sqrt{\frac{\hbar}
{2(2\pi)^3\varepsilon_0\omega_{\vec{k}}}}[a_{\vec{k}\lambda}(t)
e^{\imath\vec{k}\cdot\vec{r}}+a_{\vec{k}
\lambda}^\dag(t)e^{-\imath\vec{k}\cdot\vec{r}}]\vec{e}_{\vec{k}\lambda}
\end{equation}
where $ \omega_{\vec{k}}=c|\vec{k}| $, $\varepsilon_0 $ is the
permittivity of the vacuum and
 $\vec{e}_{\vec{k}\lambda}, \hspace{00.50 cm} (\lambda=1,2) $ are
unit polarization vectors which satisfy the orthogonality
relations
\begin{eqnarray}\label{d2}
\vec{e}_{\vec{k}\lambda}\cdot\vec{e}_{\vec{k}\lambda'}&=&
\delta_{\lambda\lambda'},\nonumber\\
 \vec{e}_{\vec{k}\lambda}\cdot\vec{k}&=&0,
\end{eqnarray}
Operators $a_{\vec{k}\lambda}(t)$ and $a_{\vec{k}\lambda}^\dag(t)$
are annihilation and creation operators of the electromagnetic
field and satisfy the following equal time commutation rules
\begin{equation}\label{d3}
[a_{\vec{k}\lambda}(t),a_{\vec{k'}\lambda'}^\dag(t)]=
\delta(\vec{k}-\vec{k'})\delta_{\lambda\lambda'}.
\end{equation}
Quantization in  coloumb gauge usually needs resolution of a
vector field in its transverse and longitudinal parts. Any vector
field $\vec{F}(\vec{r})$ can be resolved in two components,
transverse and  longitudinal components which are denoted
respectively by $ \vec{F}^\bot$ and $\vec{F}^\|$ \cite{17,18}. The
transverse part satisfies the coloumb condition $
\nabla\cdot\vec{F}^\bot=0$ and the longitudinal component is a
conservative field $ \nabla\times\vec{F}^\|=0$. In the absence of
external charges the displacement field is purely transverse and
can be expanded    in terms of the plane waves as
\begin{equation}\label{d4}
 \vec{D}(\vec{r},t)=-\imath\varepsilon_0\int d^3\vec{q}
\sum_{\lambda=1}^2
\sqrt{\frac{\hbar\omega_{\vec{k}}}{2(2\pi)^3\varepsilon_0}}
[a_{\vec{k}\lambda}^\dag(t)e^{-\imath\vec{k}\cdot\vec{r}}-a_{\vec{k}\lambda}(t)
e^{\imath\vec{k}\cdot\vec{r}}]\vec{e}_{\vec{k}\lambda}.
\end{equation}
where $\mu_0$ is the  permeability  of the vacuum.
 From (\ref{d3}) the
commutation relations between the components of the vector
potential and the displacement field clearly are
\begin{equation}\label{d5}
[A_l(\vec{r},t),-D_j(\vec{r'},t)]=
\imath\hbar\delta_{lj}^\bot(\vec{r}-\vec{r'}),
\end{equation}
where $\delta_{lj}^\bot(\vec{r}-\vec{r'})=\frac{1}{(2\pi)^3}\int
d^3\vec{k}e^{\imath\vec{k}\cdot(\vec{r}-\vec{r'})}(\delta_{lj}-\frac{k_l
k_j }{|\vec{k}|^2})$, is the transverse delta function.

Now we enter the medium in the process of quantization taking its
contribution in  Hamiltonian of the total system (
electromagnetic field plus medium) as the sum of the Hamiltonians
of the $E$ and $M$ quantum fields
\begin{eqnarray}\label{d6}
H_d&=&H_e+H_m, \nonumber\\
 H_e(t)&=&\sum_{\nu=1}^3\int d^3\vec{q}\int d^3\vec{k}
 \ \hbar\omega_{\vec{k}}
\ d_{\nu}^\dag(\vec{k},\vec{q},t)d_{\nu}(\vec{k},\vec{q},t)
,\nonumber\\
H_m(t)&=&\sum_{\nu=1}^3\int d^3\vec{q}\int d^3\vec{k} \
\hbar\omega_{\vec{k}}\
b_{\nu}^\dag(\vec{k},\vec{q},t)b_{\nu}(\vec{k},\vec{q},t) .
\end{eqnarray}
where $H_e$ and $H_m$ are the Hamiltonians the $E$ and $M$ fields
respectively. Quantum dynamics of a dissipative harmonic
oscillator interacting with an absorptive environment can be
investigated by modeling the environment by a continuum of
harmonic oscillators \cite{19}-\cite{27}. In the case of
quantization of electromagnetic field in the presence of a
magnetodielectric medium, electromagnetic field is the main
dissipative system and the medium plays the role of the absorptive
environment. Here electromagnetic field contains a continuous set
of harmonic oscillators labeled by $\vec{k}$ and $\nu=1,2$.
Therefore to each harmonic oscillator of the electromagnetic
field a continuum of oscillators should be corresponded. In the
present scheme, to each harmonic oscillator of the
electromagnetic field labeled by $\vec{k}$ and $\nu $, we have
corresponded two continuous set of harmonic oscillators defined
by the ladder operators $d_\nu(\vec{k},\vec{q}, t)$,
$d^{\dag}_\nu(\vec{k},\vec{q}, t)$ and $b_\nu(\vec{k},\vec{q},
t)$, $b^{\dag}_\nu(\vec{k},\vec{q}, t)$ which are to describe the
electric and magnetic properties of the medium respectively and
satisfy the equal time commutation relations
\begin{eqnarray}\label{d7}
&&[d_{\nu}(\vec{k},\vec{q},t) ,
d_{\nu'}^\dag(\vec{k'},\vec{q'},t)]=
\delta_{\nu\nu'}\delta(\vec{k}-\vec{k}')\delta(\vec{q}-\vec{q}'),\nonumber\\
&&[b_{\nu}(\vec{k},\vec{q},t),b_{\nu'}^\dag(\vec{k'},\vec{q'},t)]=
\delta_{\nu\nu'}\delta(\vec{k}-\vec{k}')\delta(\vec{q}-\vec{q}').
\end{eqnarray}
Summation on $\nu=3$ in (\ref{d6}) is necessary, because as we
will see the polarization densities of the medium are defined in
terms of the ladder operators of the $E$ and $M$ fields and in
opposite to the vector potential the polarization densities are
not purely transverse. In relation (\ref{d6}) $ \omega_{\vec{k}}$
is called dispersion relation of the magnetodielectric medium and
can be chosen simply as $\omega_{\vec{k}}=c|\vec{k}|$
\cite{15,16}. It is remarkable that, although the medium is
anisotropic in its electric and magnetic properties, we do not
need  take the dispersion relation as a tensor. In fact the
dispersion relation has not  any physical meaning here, and the
Hamiltonian of the medium  as (\ref{d6}) is merely a mathematical
institution so that  gives us the correct form of the equation of
motion of the total system, that is the
 Maxwell and the constitutive equations of the magnetodielectric medium.
The anisotropic behavior of the medium is expressed merely as
definitions of the electric  polarization density,
 denoted  by $\vec{P}$ and magnetic polarization density, denoted $\vec{M}$, in terms of the
ladder operators of the $E$ and $M$ fields as follows
\begin{eqnarray}\label{d8}
P_i(\vec{r},t)&=&\sum_{\nu=1}^3 \int
\frac{d^3\vec{q}}{\sqrt{(2\pi)^3}} \int d^3\vec{k}\int
d^3r'f_{ij}(\omega_{\vec{k}},\vec{r},\vec{r'})\nonumber\\
&\times &\left[d_{\nu}(\vec{k},\vec{q},t)e^{i\vec{q}\cdot\vec{r'}}+
H.c.\right] v^j_{\nu}(\vec{q}),
\end{eqnarray}
\begin{eqnarray}\label{d9}
 M_i(\vec{r},t)&=&\imath\sum_{\nu=1}^3 \int
\frac{d^3\vec{q}}{\sqrt{(2\pi)^3}}\int d^3\vec{k}\int d^3r'
g_{ij}(\omega_{\vec{k}},\vec{r} , \vec{r'})\nonumber\\
&\times&\left[b_{\nu}(\vec{k},\vec{q},t)e^{\imath\vec{q}\cdot\vec{r'}}-
H.c.\right]s^j_{\nu}(\vec{q}),
\end{eqnarray}
where
\begin{eqnarray}\label{d10}
&&\vec{v}_{\nu}(\vec{q})=\vec{e}_{\vec{q}\nu},\hspace{3cm}\nu=1,2\nonumber\\
\\
&&\vec{s}_{\nu}(\vec{q})=\hat{q}\times\vec{e}_{\nu\vec{q}}
,\hspace{1cm}\nu=1,2\nonumber\\\
\\
&&\vec{v}_{3}(\vec{q})=\vec{s}_{3}(\vec{q})=\hat{q}=
\frac{\vec{q}}{|\vec{q}|},\nonumber\\
\end{eqnarray}
In (\ref{d8}) and (\ref{d9}), the
 $ f_{ij}(\omega_{\vec{k}},\vec{r}, \vec{r'})$ and $g_{ij}(\omega_{\vec{k}},\vec{r}, \vec{r'}) $,
 are some real valued tensors and are called  the coupling
 tensors of  electromagnetic field and the medium. The coupling tensors
   play the key role in this method
  and are a measure for the strength of the polarizability and magnetizability of the
  medium macroscopically, so that we will see that the electric and magnetic
  susceptibility tensors of the magnetodielectric medium are obtained in terms of
 these  coupling tensors. Also, the explicit forms of the noise polarization densities
  are obtained in terms of the coupling tensors and the ladder operators of the
$E$ and $M$ fields at $t=0$. It can be shown that when the medium
tends  a non absorbing one, the noise densities tend to zero and
this quantization scheme reduce to the usual quantization in
these media \cite{15}.

Now  the total Hamiltonian, i.e., electromagnetic field plus the
E and M quantum fields can be proposed as one of the following
forms
\begin{eqnarray}\label{d11}
\tilde{H}(t)&=&\int d^3r \left[\frac{[
\vec{D}-\vec{P}]^2}{2\varepsilon_0}+
\frac{[\nabla\times\vec{A}-\mu_0\vec{M}]^2}{2\mu_0}\right]+H_e+H_m,\nonumber\\
&&
\end{eqnarray}
\begin{eqnarray}\label{d11.1}
\tilde{\tilde{H}}(t)&=&\int d^3r \left[\frac{[
\vec{D}-\vec{P}]^2}{2\varepsilon_0}+
\frac{[\nabla\times\vec{A}]^2}{2\mu_0}-\nabla\times\vec{A}\cdot\vec{M}\right]+H_e+H_m,\nonumber\\
&&
\end{eqnarray}
  in which the electric and magnetic polarizations interact
minimally with the displacement field and the magnetic field,
respectively. Using both of the Hamiltonians (\ref{d11}) and
(\ref{d11.1}) give us correctly the Maxwell and constitutive
equations of the manetodielectric medium as the Heisenberg
equations of the total system. Here we use the Hamiltonian
(\ref{d11}) since it is easier to solve the coupled Maxwell and
constitutive equations especially when the medium is translation
invariant in its electric and magnetic properties.
\section{Heisenberg equations}
\subsection{Maxwell equations}

The Heisenberg equations for the fields $\vec{A}$ and $\vec{D}$ are
\begin{equation}\label{d12}
\frac{\partial\vec{A}(\vec{r},t)}{\partial
t}=\frac{\imath}{\hbar}[\tilde{H},\vec{A}(\vec{r},t)]=
-\frac{\vec{D}(\vec{r},t)-\vec{P}^\bot(\vec{r},t)}{\varepsilon_0},
\end{equation}
\begin{equation}\label{d13}
\frac{\partial\vec{D}(\vec{r},t)}{\partial
t}=\frac{\imath}{\hbar}[\tilde{H},\vec{D}(\vec{r},t)]=
\frac{\nabla\times\nabla\times\vec{A}(\vec{r},t)}{\mu_0}-\nabla\times\vec{M}
(\vec{r},t),
\end{equation}
where $\vec{P}^\bot$ is the transverse component of $\vec{P}$. If
we define the transverse electric field $\vec{E}^\bot $, magnetic
induction $\vec{B}$ and magnetic field $\vec{H}$ as
 \begin{equation}\label{d14}
 \vec{E}^\bot=-\frac{\partial\vec{A}}{\partial t},\hspace{1.00
 cm}\vec{B}=\nabla\times\vec{A},\hspace{1.00
 cm}\vec{H}=\frac{\vec{B}}{\mu_0}-\vec{M}.
 \end{equation}
The Heisenberg equations  (\ref{d12}) and (\ref{d13}) can be
rewritten as
\begin{equation} \label{d15}
\vec{D}=\varepsilon_0 \vec{E}^\bot+\vec{P}^\bot,
\end{equation}
\begin{equation}\label{d16}
\frac{\partial \vec{D}}{\partial t}=\nabla\times\vec{H},
\end{equation}
which are respectively the definition of the displacement field
and the macroscopic Maxwell equation in the absence of external
charges. In the absence of external charges. In coloumb gauge  we
can take the longitudinal component of the electric field as
$\vec{E}^\|=-\frac{\vec{P}^\|}{\varepsilon_0}$. According to the
definitions  (\ref{d14}) we have
\begin{equation}\label{d16.1}
\nabla\times\vec{E}=-\frac{\partial\vec{B}}{\partial t}.
\end{equation}
\subsection{Constitutive equations of the magnetodielectric medium}

Using commutation relations (\ref{d7}) we easily find the
Heisenberg equations for operators $ d_{\nu}(\vec{k},\vec{q},t) $
and $ b_{\nu}(\vec{k},\vec{q},t)$ as
\begin{eqnarray}\label{d17}
&&\dot{d}_{\nu}(\vec{k},\vec{q},t)=
\frac{\imath}{\hbar}[\tilde{H},d_{\nu}(\vec{k},\vec{q},t)]=
-\imath\omega_{\vec{k}}\ d_{\nu}(\vec{k},\vec{q},t)\nonumber\\
&&+\frac{\imath}{\hbar\sqrt{(2\pi)^3}} \int d^3r'\int d^3r''\
e^{-\imath\vec{q}\cdot
\vec{r}''}f_{ij}(\omega_{\vec{k}},\vec{r'},\vec{r''})\
E^i(\vec{r'},t)\ v^j_{\nu}(\vec{q}),\nonumber\\
&&
\end{eqnarray}
\begin{eqnarray}\label{d18}
&&\dot{b}_{\nu}(\vec{k},\vec{q},t)=\frac{\imath}{\hbar}[\tilde{H},b_{\nu}(\vec{k},\vec{q},t)]=
 -\imath\omega_{\vec{k}}\ b_{\nu}(\vec{k},\vec{q},t)\nonumber\\
&&+\frac{\mu_0}{\hbar\sqrt{(2\pi)^3}} \int d^3r'\int d^3r'' \
e^{-\imath\vec{q}\cdot\vec{r}''}
g_{ij}(\omega_{\vec{k}},\vec{r'},\vec{r}'')\ H^i(\vec{r'},t)
\ s^j_{\nu}(\vec{q}).\nonumber\\
&&
\end{eqnarray}
These equations can be solved formally  as
\begin{eqnarray}\label{d19}
&&{d}_{\nu}(\vec{k},\vec{q},t)=
d_{\nu}(\vec{k},\vec{q},0)e^{-\imath\omega_{\vec{k}}t}+
\frac{\imath}{\hbar\sqrt{(2\pi)^3}}\int_0^t
dt'\ e^{-\imath\omega_{\vec{k}}(t-t')}\nonumber\\
&&\times \int d^3r'\int d^3r'' \
e^{-i\vec{q}\cdot\vec{r}''}f_{ij}(\omega_{\vec{k}},\vec{r'},\vec{r}'')
E^i(\vec{r'},t')v^j_{\nu}(\vec{q}),\nonumber\\
&&
\end{eqnarray}
\begin{eqnarray}\label{d20}
&&{b}_{\nu}(\vec{k},\vec{q},t)=
b_{\nu}(\vec{k},\vec{q},0)e^{-\imath\omega_{\vec{k}}t}+
\frac{\mu_0}{\hbar\sqrt{(2\pi)^3}}\int_0^t
dt'\ e^{-\imath\omega_{\vec{k}}(t-t')}\nonumber\\
&&\times \int d^3r'\int d^3r''\
e^{-i\vec{q}\cdot\vec{r}''}g_{ij}(\omega_{\vec{k}},\vec{r'},\vec{r}'')H^i
(\vec{r'},t')s^j_{\nu}(\vec{q}).\nonumber\\
&&
\end{eqnarray}
 Substituting of (\ref{d19}) in (\ref{d8})  and (\ref{d20}) in (\ref{d9})
 and using the completeness relations
\begin{equation}\label{d20.1}
\sum_{\nu=1}^3e^i_{\nu\vec{q}}e^j_{\nu\vec{q}}=\sum_{\nu=1}^3s^i_{\nu\vec{q}}
s^j_{\nu\vec{q}}=\delta_{ij},
\end{equation}
give us the constitutive equations of the magnetodielectric medium
 which relate the electric and magnetic polarization densities
of the medium to the macroscopic electric and magnetic fields
respectively as
\begin{equation}\label{d21}
P_i(\vec{r},t)=P_{Ni}(\vec{r},t)+\varepsilon_0\int_0^{|t|} d
t'\int d^3r' \
\chi^e_{ij}(\vec{r},\vec{r'},|t|-t')E^j(\vec{r'},\pm t'),
\end{equation}
\begin{equation}\label{d22}
M_i(\vec{r},t)=M_{Ni}(\vec{r},t)+\int_0^{|t|} dt'\int d^3r' \
\chi^m_{ij}(\vec{r},\vec{r'},|t|-t')H^j(\vec{r'},\pm t'),
\end{equation}
where the upper (lower) sign corresponds to $t>0$ ($ t<0 $) and
$\vec{E}=-\frac{\partial\vec{A}}{\partial
t}-\frac{\vec{P}^\|}{\varepsilon_0}$, is the total electric
field. The memory tensors
\begin{eqnarray}\label{d23}
&&\chi^e(\vec{r}, \vec{r'}, t)=\nonumber\\
&&\nonumber\\
 &&\left\{\begin{array}{cc}
 \displaystyle \frac{8\pi}{\hbar c^3 \varepsilon_0}\int_0^\infty
d\omega\omega^2\sin\omega t\int d^3r''
\left[\ f(\omega,\ \vec{r}\ ,\ \vec{r''})\ \cdot \ f^t(\omega\ ,\vec{r'}\ ,\ \vec{r''})\right] & \hspace{1.00cm} t>0 \\
\\
\\
0 & \hspace{1.00 cm}t\leq0
\end{array}\right.\nonumber\\
\nonumber\\
&&
\end{eqnarray}
and
\begin{eqnarray}\label{d24}
&&\chi^m(\vec{r}, \vec{r'} , t)=\nonumber\\
&&\nonumber\\
 &&\left\{\begin{array}{cc}
  \displaystyle \frac{8\pi\mu_0}{\hbar c^3 }\int_0^\infty
d\omega\omega^2\sin\omega t\int d^3r''
\left[\ g(\omega,\ \vec{r}\ ,\ \vec{r''})\ \cdot \ g^t(\omega\ ,\vec{r'}\ ,\ \vec{r''})\right]  & \hspace{1.00 cm}  t>0 \\
\\
\\
0 & \hspace{1.00 cm}t\leq0
\end{array}\right.\nonumber\\
\nonumber\\
&&
\end{eqnarray}
 are respectively the electric and magnetic susceptibility tensors of the
magnetodielectric medium which have been obtained in terms of the
coupling tensors $ f$ , $g$  and their transpositions $ f^t$ ,
$g^t$. For a   medium with a definite pair of the tensors $
\chi^e$ and $\chi^m$  it is possible to solve equations
(\ref{d23}) and (\ref{d24}) in terms of the coupling tensors $ f$
and $g$  using a type of eigenvalue problem \cite{14.1} . The
operators $\vec{P}_N $ and $\vec{M}_N$ in (\ref{d21}) and
(\ref{d22}) are the noise electric and magnetic polarization
densities and  their explicit forms are given by
\begin{eqnarray}\label{d28}
P_{Ni}(\vec{r},t)&=&
\sum_{\nu=1}^3\int\frac{d^3\vec{q}}{\sqrt{(2\pi)^3}}\int
d^3\vec{k}\int d^3r'\
f_{ij}(\omega_{\vec{k}},\vec{r},\vec{r'})\nonumber\\
 &\times &\left[d_{\nu}(\vec{k},\vec{q},0)
e^{-\imath\omega_{\vec{k}}t+\imath\vec{q}\cdot\vec{r'}}+H.c.
\right]v^j_{ \nu}(\vec{q}),
\end{eqnarray}
\begin{eqnarray}\label{d29}
M_{Ni}(\vec{r},t)&=&
i\sum_{\nu=1}^3\int\frac{d^3\vec{q}}{\sqrt{(2\pi)^3}} \int
d^3\vec{k}\int d^3r'\
g_{ij}(\omega_{\vec{k}},\vec{r},\vec{r'})\nonumber\\
&\times &\left[b_{\nu}(\vec{k},\vec{q},0)
e^{-\imath\omega_{\vec{k}}t+\imath\vec{q}\cdot\vec{r'}}-H.c.\right]
s^j_{\nu}(\vec{q}).
\end{eqnarray}
From (\ref{d23}) and (\ref{d24}) it is clear that for a given pair
of the susceptibility tensors $\chi^e$ and $\chi^m$, the
solutions of the relations (\ref{d23})and (\ref{d24}) for  the
coupling tensors $f$ and $g$  are not unique. In fact for a given
pair of $\chi^e$ and $\chi^m$, if the tensors $f$ and $g$ satisfy
equations (\ref{d23}) and (\ref{d24}), then the coupling tensors
\begin{eqnarray}\label{d29.1}
&& f'( \omega , \vec{r}, \vec{r'})=\int \ d^3s\ f( \omega ,
\vec{r} , \vec{s})\ \cdot\  A^t( \omega ,
\vec{s},\vec{r'})\nonumber\\
&& g'( \omega , \vec{r}, \vec{r'})=\int \ d^3s\ g( \omega ,
\vec{r} , \vec{s})\ \cdot\  A^t( \omega , \vec{s},\vec{r'})
\end{eqnarray}
for any tensor $A$ where satisfy the orthogonality relation
 \begin{equation}\label{29.2}
 \int \ d^3r'' \ A( \omega , \vec{s} , \vec{r''}) \ \cdot \
A^t( \omega , \vec{s'},\vec{r''})=I\delta( \vec{s}-\vec{s'})
\end{equation}
also satisfy equations (\ref{d23}) and  (\ref{d24}). Although this
affect the space-time dependence of the noise polarizations and
therefore the space-time dependence of the electromagnetic field
operators but all of these are equivalent. This means that
various choices of the coupling tensors $f$ and $g$ satisfying
(\ref{d23}) and (\ref{d24}), for a given pair of the
susceptibility tensors $\chi^e$ and $\chi^m$, do not affect the
commutation relations between the electromagnetic field operators
and hence the physical observables. This subject becomes more
clear if we compute the commutation relations between the
components of the temporal  Fourier transformations of the noise
polarizations $\vec{P}_N$ and $\vec{M}_N$ using commutation
relations (\ref{d7}) and obtain
\begin{eqnarray}\label{d30}
&&[\hat{P}_{Ni}(\vec{r},\omega)\  ,\
\hat{P}_{Nj}^\dag(\vec{r'},\omega')]=\frac{\hbar\varepsilon_0}{\pi}Im\left[
\hat{\chi}^e_{ij}(\vec{r},\vec{r'},\omega)\right]\delta(\omega-\omega')\nonumber\\
&&[\hat{M}_{Ni}(\vec{r},\omega) \ , \
\hat{M}_{Nj}^\dag(\vec{r'},\omega')]=\frac{\hbar}{\mu_0
\pi}Im\left[
\hat{\chi}^m_{ij}(\vec{r},\vec{r'},\omega)\right]\delta(\omega-\omega')
\end{eqnarray}
where $\hat{\chi}^e$ and $\hat{\chi}^m $ are respectively the
temporal fourier transformations of the tensors $\chi^e $ and $
\chi^m $. The commutation relations (\ref{d30}) are
generalization of those in reference \cite{8} for an anisotropic
magnetodielectric medium with spatial-temporal dispersion.
 For a given pair of the tensors
$\chi^e$ and $ \chi^m$ , various choices of the coupling tensors
$f$ and $g$ satisfying the relations
  (\ref{d23}) and (\ref{d24}), do not affect the commutation
  relations (\ref{d30}) and therefore the commutation relations between the
  electromagnetic field operators. Therefore all of the field
  operators which are obtained with a definite pair of the
  susceptibility tensors
  $\chi^e $ and $\chi^m$ but with different coupling tensors satisfying
  (\ref{d23}) and (\ref{d24}), are equivalent.

It is clear from (\ref{d28}) and (\ref{d29}) that explicit forms
of the noise polarization densities are known. Also, because the
coupling functions $ f$, $g $ are common factors in the noise
densities $ \vec{P}_N$, $\vec{M}_N $ and the susceptibility
tensors $\chi^e$, $\chi^m $, it is clear that the strength of the
noise fields are dependent on the strength of $\chi^e$ and
$\chi^m $ which describe dissipative character of the
magnetodielectric medium. When the medium tends to a nonabsorbing
one the noise polarization tends to zero and this quantization
method is reduced to the quantization in the presence of a
nonabsorbing medium \cite{15}.

It should be noted  that the time derivative of the polarization
fields $ \frac{\partial\vec{P}}{\partial t}$
 and $ \frac{\partial\vec{M}}{\partial t}$ are continuous at time $t=0$ although
the absolute value
 $|t|$ is appeared in the constitutive equations (\ref{d21}) and (\ref{d22}). In fact  an other solution
 for the Heisenberg equation (\ref{d17}) can be written as
 \begin{eqnarray}\label{d30.1}
 &&{d}_{\nu}(\vec{k},\vec{q},t)=
d^{in}_{\nu}(\vec{k},\vec{q})e^{-\imath\omega_{\vec{k}}t}+
\frac{\imath}{\hbar\sqrt{(2\pi)^3}}\int_{-\infty}^t
dt'e^{-\imath\omega_{\vec{k}}(t-t')} \nonumber\\
&&\times \int d^3r'\int d^3r''
e^{-i\vec{q}\cdot\vec{r}''}f_{ij}(\omega_{\vec{k}},\vec{r'},
\vec{r}'') E^i(\vec{r'},t')v^j_{\nu}(\vec{q})
\end{eqnarray}
where $d^{in}_{\nu}(\vec{k},\vec{q})$  are some time independent
annihilation operators which satisfy the same commutation
relations (\ref{d7}). If we find  ${d}_{\nu}(\vec{k},\vec{q},0)$
from the equation (\ref{d30.1}) and  substitute it in the
 noise polarization field $\vec{P}_N(\vec{r},t)$ given by (\ref{d28}), we deduce
 \begin{eqnarray}\label{d30.2}
&&P_{Ni}(\vec{r},t)=
P^{in}_{Ni}(\vec{r},t)+\varepsilon_0\int_{-\infty}^{t} d t'\int
d^3r'\chi^e_{ij}(\vec{r},\vec{r'},t-t')E^j(\vec{r'},t')\nonumber\\
&& -\varepsilon_0\int_{0}^{|t|} d t'\int d^3r'
\chi^e_{ij}(\vec{r},\vec{r'},|t|-t')E^j(\vec{r'},\pm t'),
\end{eqnarray}
where the susceptibility tensor $\chi^e$ is given by (\ref{d23})
and $ \vec{P}^{in}_{N} $ is the same as $\vec{P}_N$  with the
exception that the annihilation operators $
d_\nu(\vec{k},\vec{q},0)$ should be replaced by $
d^{in}_\nu(\vec{k},\vec{q})$. Now sustituting  $\vec{P}_N$ from
(\ref{d30.2}) into (\ref{d21}) give us the expression
\begin{eqnarray}\label{d30.3}
&&P_i(\vec{r},t)= P^{in}_{Ni}(\vec{r},t)
+\varepsilon_0\displaystyle\int_{-\infty}^{t} d t'\int d^3r'
\chi^e_{ij}(\vec{r},\vec{r'},t-t')E^j(\vec{r'}, t')\nonumber\\
&&
\end{eqnarray}
 for the polarization field $\vec{P}$ which is valid for both
positive and negative times. From (\ref{d30.3}) it is clear that
$\frac{\partial \vec{P}}{\partial t}$ and accordingly the
electromagnetic field operators are continuous at $t=0$. One can
apply the constitutive equation (\ref{d30.3}) and similar equation
for the magnetic polarization $\vec{M}$ and use the temporal
Fourier transformation, or apply the constitutive equations
(\ref{d21}) and (\ref{d22}) and use the Laplace transformation, to
solve the coupled constitutive and Maxwell equations. Here we
prefer the later, since in this way it is easier  to show the
limiting cases in the absence of any medium or in the presence of
a nonabsorbing medium \cite{15,16}. Using the forward and backward
Laplace transformation and applying the constitutive equations
(\ref{d21}) and (\ref{d22}) , we can obtain the explicit forms of
the electromagnetic operators for both negative and positive
times which are continuous at $t=0$\cite{12}.
\section{Solution of the Heisenberg equation}

In this section we solve  the coupled Maxwell and constitutive
equations (\ref{d15}),(\ref{d16}), (\ref{d16.1}), (\ref{d21}) and
(\ref{d22})  for a translationally  invariant medium, that is for
a medium that its electric and magnetic susceptibility tensors
 are dependent on the difference $\vec{r}-\vec{r'}$. For such a medium we can obtain the explicit
forms of the electromagnetic field operators easily for both
positive and negative times by the spatial fourier transformation
and the temporal Laplace transformation. In this section we solve
the coupled Maxwell and constitutive equations for positive times
using  the forward Laplace transformation.  The solution of the
Heisenberg equations for negative times can be found similarly
from the backward Laplace transformation \cite{12}. Let us define
$\underline{\hat{\vec{F}}}$ for any vector field $F(\vec{r},t)$ by
\begin{equation}\label{d32}
\underline{\hat{\vec{F}}}(\vec{k},\rho)=\int
d^3r\int_{0}^{\infty}dt\ \vec{F}(\vec{r},t)\
e^{-\imath\vec{k}\cdot\vec{r}-\rho t}.
\end{equation}
Now applying such a transformation on the both sides of  the
 equations the constitutive
equations (\ref{d15}), (\ref{d21}) and (\ref{d22}), together with
the Maxwell equations (\ref{d16}) and (\ref{d16.1}), and then
their combination  for a translationally invariant medium we find
\begin{eqnarray}\label{d33}
&&-\imath\vec{k}\times\underline{\hat{\vec{E}}}=-\rho\underline{\hat{\mu}}(\vec{k},p)
\underline{\hat{\vec{H}}}-\mu_0\rho\underline{\hat{\vec{M}}}_N(\vec{k},\rho)+
\underline{\vec{B}}(\vec{k},0),\nonumber\\
&&-\imath\vec{k}\times\underline{\hat{\vec{H}}}=\rho\underline{\hat{\varepsilon}}
(\vec{k},\rho)\underline{\hat{\vec{E}}}+\rho\underline{\hat{\vec{P}}}_N(\vec{k},\rho)-
\underline{\vec{D}}(\vec{k},0),
\end{eqnarray}
where $\underline{\vec{B}}(\vec{k},0)$ ,
$\underline{\vec{D}}(\vec{k},0)$ are respectively the spatial
fourier transformations of $\vec{B}(\vec{r},0)$ and
$\vec{D}(\vec{r},0)$ and $
\underline{\hat{\varepsilon}}=\varepsilon_0(1+\underline{\hat{\chi}}^e)$
, $\underline{\hat{\mu}}=\mu_0(1+\underline{\hat{\chi}}^m)$, are
the transformation introduced in (\ref{d32}) for permittivity and
permeability  tensors of the medium.  The equations (\ref{d33})
can be written in a compact form using a matrix notation as
follows
\begin{equation}\label{d34}
\Lambda (\vec{k},\rho)\left[ \begin{array}{c}
  \underline{\hat{\vec{E}}} \\
  \\
  \underline{\hat{\vec{H}}}
\end{array}\right]
=\left[\begin{array}{c}
  \mu_0\rho\underline{\hat{\vec{M}}}_N-\underline{\vec{B}}(\vec{k},0) \\
\\
 - \rho\underline{\hat{\vec{P}}}_N+\underline{\vec{D}}(\vec{k},0)
\end{array}\right],
\end{equation}
where $ \Lambda(\vec{k},\rho)$ is a $6\times 6$ matrix defined by
\begin{equation}\label{d35}
\Lambda(\vec{k},\rho)=\left[\begin{array}{cc}
  O(\vec{k}) & -\rho\underline{\hat{\mu}}(\vec{k},\rho) \\
  \\
  \rho\underline{\hat{\varepsilon}}(\vec{k},\rho) & O(\vec{k})
\end{array}\right],
\end{equation}
and
\begin{equation}\label{d36}
O(\vec{k})=\left[\begin{array}{ccc}
  0 & -\imath k_3 & \imath k_2 \\
  \\
  \imath k_3 & 0 & -\imath k_1 \\
  \\
  -\imath k_2 & \imath k_1 & 0
\end{array}\right].
\end{equation}
Finally,  substituting  $
\underline{\hat{\vec{P}}}_N(\vec{k},\rho) $,
$\underline{\hat{\vec{M}}}_N(\vec{k},\rho)$, $\vec{D}(\vec{k},0)$
and $\vec{B}(\vec{k},0)$ in the right hand side of (\ref{d34})
using the expansions (\ref{d1}), (\ref{d4}), (\ref{d28}) and
(\ref{d29}) and multiplying the equation (\ref{d34}) on the left
 by $\Lambda^{-1}(\vec{k},\rho)$, we find
\begin{eqnarray}\label{d37}
E_i(\vec{r},t)&=&\imath \sum_{ \lambda=1}^2\int
d^3k\sqrt{\frac{\hbar\omega_{\vec{k}}\varepsilon_0}{2(2\pi)^3}}\left[
\gamma_{ij}(\vec{k},t)a_{\vec{k}\lambda}(0)e^{\imath\vec{k}\cdot\vec{r}}-H.c.\right]
e^j_{\vec{k}\lambda}\nonumber\\
&+&\imath \sum_{ \lambda=1}^2\int
d^3k\sqrt{\frac{\hbar\omega_{\vec{k}}\mu_0}{2(2\pi)^3}}\left[
\xi_{ij}(\vec{k},t)a_{\vec{k}\lambda}(0)e^{\imath\vec{k}\cdot\vec{r}}-H.c.\right]
s^j_{\vec{k}\lambda}\nonumber\\
&+& \imath\sum_{\nu=1}^3\int d^3q \int
\frac{d^3k}{\sqrt{(2\pi)^3}}\left[
\zeta_{ij}(\omega_{\vec{q}},\vec{k},t)b_\nu(\vec{q},\vec{k},0)
e^{\imath\vec{k}\cdot\vec{r}}-H.c.\right]s^j_{\vec{k}\nu}\nonumber\\
&+& \sum_{\nu=1}^3\int d^3q \int
\frac{d^3k}{\sqrt{(2\pi)^3}}\left[
\eta_{ij}(\omega_{\vec{q}},\vec{k},t)d_\nu(\vec{q},\vec{k},0)
e^{\imath\vec{k}\cdot\vec{r}}+H.c.\right]e^j_{\vec{k}\nu},\nonumber\\
&&
\end{eqnarray}
 where
$\omega_{\vec{k}}=c|\vec{k}|$, $\omega_{\vec{q}}=c|\vec{q}|$ and we
have used $\Lambda(-\vec{k},\rho)=\Lambda^*(\vec{k},\rho)$. The
tensors $\gamma$, $\xi$, $\zeta$ and $\eta $ are given by
\begin{eqnarray}\label{d38}
\gamma_{ij}(\vec{k},t)&=&L^{-1}\left[\left[
\Lambda(\vec{k},\rho)\right]^{-1}_{i(j+3)}\right],\nonumber\\
\xi_{ij}(\vec{k},t)&=&-L^{-1}\left[\left[
\Lambda(\vec{k},\rho)\right]^{-1}_{ij}\right],\nonumber\\
\zeta_{ij}(\omega_{\vec{q}},\vec{k},t)&=&\mu_0L^{-1}\left[\frac{\rho}
{\rho+\imath\omega_{\vec{q}}} \sum_{l=1}^3\left(\ \left[
\Lambda(\vec{k},\rho)\right]^{-1}_{il}\ \underline{g}_{lj}
( \omega_{\vec{q}},\vec{k})\right)\right],\nonumber\\
\eta_{ij}(\omega_{\vec{q}},\vec{k},t)&=&-L^{-1}\left[\frac{\rho}
{\rho+\imath\omega_{\vec{q}}} \sum_{l=1}^3\left(\
\left[\Lambda(\vec{k},\rho)\right]^{-1}_{i(l+3)}\
\underline{f}_{lj}( \omega_{\vec{q}},\vec{k})\right)\right],\nonumber\\
&&
\end{eqnarray}
where $ L^{-1}\left\{f(\rho)\right\}$ denotes the inverse Laplace
transform of $ f(\rho)$ and the tensors $ \underline{f}$,
$\underline{g}$ are given  by
\begin{equation}\label{d38.1}
\underline{f}_{ij}(\omega , \vec{k})=\int d^3r f_{ij}( \omega ,
\vec{r}) e^{-\imath\vec{k}\cdot\vec{r}}\hspace{2.00
cm}\underline{g}_{ij}(\omega , \vec{k})=\int d^3r g_{ij}( \omega ,
\vec{r}) e^{-\imath\vec{k}\cdot\vec{r}}
\end{equation}
 Similarly for the magnetic field $\vec{H}$ we obtain
\begin{eqnarray}\label{d39}
H_i(\vec{r},t)&=&\imath \sum_{ \lambda=1}^2\int
d^3k\sqrt{\frac{\hbar\omega_{\vec{k}}\varepsilon_0}{2(2\pi)^3}}\left[
\tilde{\gamma}_{ij}(\vec{k},t)a_{\vec{k}\lambda}(0)
e^{\imath\vec{k}\cdot\vec{r}}-H.c.\right]e^j_{\vec{k}\lambda}\nonumber\\
&+&\imath \sum_{ \lambda=1}^2\int
d^3k\sqrt{\frac{\hbar\omega_{\vec{k}}\mu_0}{2(2\pi)^3}}\left[
\tilde{\xi}_{ij}(\vec{k},t)a_{\vec{k}\lambda}(0)e^{\imath\vec{k}
\cdot\vec{r}}-H.c.\right]s^j_{\vec{k}\lambda}\nonumber\\
&+& \imath\sum_{\nu=1}^3\int d^3q \int
\frac{d^3k}{\sqrt{(2\pi)^3}}\left[
\tilde{\zeta}_{ij}(\omega_{\vec{q}},\vec{k},t)b_\nu(\vec{q},\vec{k},0)
e^{\imath\vec{k}\cdot\vec{r}}-H.c.\right]s^j_{\vec{k}\nu}\nonumber\\
&+& \sum_{\nu=1}^3\int d^3q \int
\frac{d^3k}{\sqrt{(2\pi)^3}}\left[
\tilde{\eta}_{ij}(\omega_{\vec{q}},\vec{k},t)d_\nu(\vec{q},\vec{k},0)
e^{\imath\vec{k}\cdot\vec{r}}+H.c.\right]e^j_{\vec{k}\nu},\nonumber\\
&&
\end{eqnarray}
where
\begin{eqnarray}\label{d40}
\tilde{\gamma}_{ij}(\vec{k},t)&=&L^{-1}\left[\left[
\Lambda(\vec{k},\rho)\right]^{-1}_{(i+3)(j+3)}\right],\nonumber\\
\tilde{\xi}_{ij}(\vec{k},t)&=&-L^{-1}\left[\left[
\Lambda(\vec{k},\rho)\right]^{-1}_{(i+3)j}\right],\nonumber\\
\tilde{\zeta}_{ij}(\omega_{\vec{q}},\vec{k},t)&=&\mu_0L^{-1}\left[\frac{\rho}
{\rho+\imath\omega_{\vec{q}}} \sum_{l=1}^3\left(\
\left[\Lambda(\vec{k},\rho)\right]^{-1}_{(i+3)l}\
\underline{g}_{lj}
(\omega_{\vec{q}},\vec{k})\right)\right],\nonumber\\
\tilde{\eta}_{ij}(\omega_{\vec{q}},\vec{k},t)&=&-L^{-1}\left[\frac{\rho}
{\rho+\imath\omega_{\vec{q}}}\sum_{l=1}^3\left(\
\left[\Lambda(\vec{k},\rho) \right]^{-1}_{(i+3)(l+3)}\
\underline{f}_{lj}( \omega_{\vec{q}},\vec{k})
\right)\right].\nonumber\\
&&
\end{eqnarray}
The expressions (\ref{d37})-(\ref{d40}) and the commutation
relations (\ref{d3}) and (\ref{d7}) show that the commutation
relations between components of the operators $\vec{E}$ and
$\vec{H}$ are independent of the various choices of the coupling
tensors $f$ and $g$ satisfying the relations (\ref{d23}) and
(\ref{d24}) for a given pair of the tensors $\chi^e$ and $\chi^m$.
Finally having the fields $\vec{E}$ and $\vec{H}$ we can obtain
the polarization fields $\vec{P}$ and $\vec{M}$ from the
constitutive equations (\ref{d21}) and (\ref{d22}).\\

It should be pointed out that hereunto we have assumed the medium
covered  by this quantization scheme is a non-conductor
polarizable and magnetizable medium.  In the case of a
non-conductor magnetodielectric medium the operator $\vec{P}$
defined by (\ref{d8}) is  the electric polarization density and
$\frac {\partial\vec{P}}{\partial t}$, as the current density
induced in the medium due to the electric polarization. When we
are concerned  with a conductor magnetodielectric  medium the
fields $\vec{P}$ and $\vec{D}$ in (\ref{d15}) may not be
interpreted as the electric polarization and displacement field
and the quantity $\frac {\partial\vec{P}}{\partial t}$ in Maxwell
equation (\ref{d16}), where can be rewritten as
\begin{equation}\label{d41}
\varepsilon_0\frac{\partial\vec{E}}{\partial t}+\frac{\partial
\vec{P}}{\partial t}=\nabla\times\vec{H},
\end{equation}
is not merely the current source created by the electric
polarization field but is the sum of the free current density ,
due to the motion of the free charges, together with the current
due to the polarizability of the medium. In this case if we
compute $ \dot{d}_\nu(\vec{k},\vec{q},t)$ from (\ref{d19}) and
substitute it in (\ref{d8}), instead of the constitutive equation
(\ref{d21}), we obtain  the  linear response relation
\begin{equation}\label{d42}
\frac{\partial P_i(\vec{r},t)}{\partial t }=
J_{Ni}(\vec{r},t)\pm\int_0^{|t|} d t'\int d^3r'\
Q_{ij}(\vec{r},\vec{r'},|t|-t')\ E^j(\vec{r'}, \pm t'),
\end{equation}
 where now
\begin{eqnarray}\label{d43}
&&Q(\vec{r}, \vec{r'}, t)=\nonumber\\
&&\nonumber\\
 &&\left\{\begin{array}{cc}
\displaystyle  \frac{8\pi}{\hbar c^3}\int_0^\infty\ d\omega\
\omega^3\cos\omega t \int d^3r''
\left[f(\omega,\ \vec{r}\ ,\ \vec{r''})\ \cdot\ f^t(\omega\ ,\vec{r'}\ ,\ \vec{r''})\right]  & t>0 \\
\\
\\
0 & t\leq0
\end{array}\right.\nonumber\\
\nonumber\\
&&
\end{eqnarray}
is $ \varepsilon_0 \frac{\partial
\chi^e(\vec{r},\vec{r'},t)}{\partial t}+\sigma
(\vec{r},\vec{r'},t)$ with $ \chi^e $ and $\sigma $ respectively
the electric susceptibility and conductivity tensors of the
medium and
\begin{eqnarray}\label{d44}
J_{Ni}(\vec{r},t)=\frac{\partial P_{Ni}}{\partial t}&=&
-\imath\sum_{\nu=1}^3\int\frac{d^3\vec{q}}{\sqrt{(2\pi)^3}}\int \
d^3\vec{k}\ \int d^3r'
\omega_{\vec{k}}\ f_{ij}(\omega_{\vec{k}},\vec{r},\vec{r'})\ \nonumber\\
 &\times &\left[d_{\nu}(\vec{k},\vec{q},0)
e^{-\imath\omega_{\vec{k}}t+i\vec{q}.\vec{r'}}-h.c \right]v^j_{
\nu}(\vec{q}),
\end{eqnarray}
is a noise current density. In this case the explicit forms of the
electromagnetic field clearly is the same as the relations
(\ref{d37})-(\ref{d40}) except that we must replace
$\rho\underline{\hat{\varepsilon}}(\vec{k},\rho)$ by
$\rho\underline{\hat{\varepsilon}}(\vec{k},\rho)+\underline{\hat{\sigma}}(\vec{k},\rho)$
where  $\underline{\hat{\sigma}}$ is the transformation introduced
in (\ref{d32}) for the conductivity tensor $\sigma$.
\section{Summary}
By modeling an anisotropic magnetodielectric medium with two
independent quantum fields, namely E and M quantum fields, we could
investigate electromagnetic field quantization in the presence of
such a medium consistently. For a given pair of the electric and
magnetic susceptibility tensors $\chi^e$ and $\chi^m$ of the medium,
we could find the corresponding coupling tensors $f$ and $g$, which
couple electromagnetic field to E and M quantum fields respectively.
The explicit space-time dependence of the noise polarizations were
obtained in terms of the ladder operators of the medium and the
coupling tensors as a consequence of Heisenberg equations. In this
approach, both the Maxwell and constitutive equations were obtained
 as Heisenberg equations of the total system. In the
limiting case, i.e., when there is no medium, this approach tends to
the usual quantization of the electromagnetic field in free space.
Also when the medium approaches to a nonabsorptive one, the noise
polarizations tend to zero and this quantization scheme is reduced
to the known quantization in such a medium.

\end{document}